\documentclass{emulateapj}

\usepackage{graphicx}
\usepackage{subfigure}

\usepackage{dcolumn}

\usepackage{color} 

\def\msun{M_{\odot}}

\def\be{\begin{equation}}

\def\ee{\end{equation}}

\def\bh{\,M_{\bullet}}
\def\kms{{\rm \,km\,s}^{-1}}
\def\yr{{\rm \,yr}}
\def\AU{{\rm \,AU}}

\def\microas{{\,\mu\rm as}}

\begin{document}

\title{Prospects for constraining the spin of the massive black hole at the
Galactic center via the relativistic motion of a surrounding star
}

\author{Qingjuan Yu$^{1,\dagger}$, Fupeng Zhang$^{1,3}$, and Youjun Lu$^{2,4}$}

\affil{$^1$\,Kavli Institute for Astronomy and Astrophysics, Peking
University, Beijing, 100871, China; $^{\dagger}$\,yuqj@pku.edu.cn\\
$^2$\,National Astronomical Observatories, Chinese Academy of
Sciences, Beijing, 100012, China; luyj@nao.cas.cn\\
$^3$\,School of Physics and Astronomy, Sun Yat-Sen University,
Guangzhou 510275, China; zhangfp7@mail.sysu.edu.cn \\
$^4$\,College of Astronomy and Space Sciences, University of Chinese Academy
of Sciences, Beijing, 100049, China
}

\begin{abstract}

In this paper, we investigate the quality of constraining the spin of the
massive black hole (MBH) at the Galactic center (GC) by using full general
relativistic simulations of the motion of a surrounding star. We obtain the
dependence mapping of the spin-induced signals on any spin direction of
the MBH for given example stars, which indicates the feasibility to test
whether the spin direction is the same as the normal of the young stellar disk
located at the GC, and, further to provide insights into the assembly history of
the MBH. We demonstrate the quality of constraining the MBH spin that may be
achieved, given any set of the astrometric and the redshift precisions of
observational facilities. We find that in the ranges of the astrometric and the
velocity precisions with 1--30$\mu$as and 0.1--10$\kms$, an improvement in
astrometric precision would be more effective at improving the quality of
constraining the spin than an improvement in velocity precision. We obtain the
parameter space of the semimajor axis and the eccentricity for the orbit of the
target star that a high-precision constraint on the GC MBH spin can be obtained
via the motion of the star.  Our results show that the spin of the GC MBH can
be constrained with 1-$\sigma$ error $\la0.1$ or even $\la0.02$ by monitoring
the orbital motion of a star, if existing as expected, with semimajor axis
$\la300\AU$ and eccentricity $\ga0.95$ over a period shorter than a decade
through future facilities.

\end{abstract}


\maketitle

\section{Introduction} 

Astronomical black holes (BHs) are believed to be described by the Kerr metric
with only two parameters, i.e., mass and spin (e.g., \citealt{Kerr63}).
Observations have exclusively demonstrated the existence of a massive BH (MBH),
with mass $\sim4\times10^6\msun$, in the Galactic center
(GC;\,\citealt{Ghezetal08,Gillessenetal09}).  However, it is still under
inspection on whether or not such an object is a Kerr BH. 

Accurately monitoring the motion of a star close to the GC MBH has long been
anticipated to provide strong dynamical constraints on the MBH spin and metric
\citep{J98,FM00,RE01,Z06,W08,KS09,AS10b,A10a,AS11,M10,SW11,Iorio11}.
Outstanding questions related to this endeavor are as follows. (1) How accurate
do the astrometric and velocity/redshift measurements need to be to provide a
tight
constraint on the MBH spin by using the motion of a star, e.g., S2/S0-2, the
closest one to the MBH currently detected \citep{Ghezetal08,Gillessenetal09}?
(2) What is the parameter space for a possibly existing star within the
orbit of S2/S0-2 that may help to put a strong constraint on the MBH spin and
metric with the next-generation (or future) facilities?  Constraining the metric
and the spin of the GC MBH is of fundamental importance for testing general
relativity (GR), and a constraint on the spin direction would also provide
insights into the assembly history of the MBH.\footnote{The spin
direction may help to reveal the assembly history of the GC MBH, which can
be understood as follows. If the MBH spun up by continuous disk accretion with
the same orientation as that of the young stellar disk at the GC, the spin
direction should align with the normal of the disk; and if the MBH grew up
via many episodes of disk accretion with random orientations, then the spin
value may be small and the spin direction may be significantly different from
that of the normal of the young stellar disk (as inferred from \citealt{KP06}).} 
In this paper, we address the above two questions and obtain guiding maps for
future observational endeavors on constraining the GC MBH spin via extensive
numerical simulations of the relativistic motion of a surrounding star in a
large parameter space. We adopt the full GR numerical method developed in
\citet{ZLY15} to investigate the spin-induced relativistic effects on the
trajectory and the redshift of a star measured by a distant observer. As an
alternative approach to those previous ones that utilize the perturbative
(post-Newtonian or weak-field) approximations \citep[e.g.,][]{W08, A10a,
AS10b}, the full GR method is efficient and accurate in obtaining the orbital
motion of a target star rotating around the GC MBH and the photon propagation
from the star to the observer. 

Whether or not those spin-induced effects can be measured and a dynamical constraint
on the GC MBH spin can be made depends on how precise the position and velocity
measurements that current and future facilities, such as the GRAVITY on the
Very Large Telescope Interferometer (VLTI), the Thirty Meter Telescope (TMT),
and the European Extremely Large Telescope (E-ELT), can achieve. For example,
for the TMT, the position precision could reach $10\mu$as for bright stars with
K-band magnitude $m_K<15$; while it could be $>50\mu$as for those stars with
$m_K>16$ because of `source confusion' \citep{2013aoel.confE..83Y}. For the
GRAVITY, the position precision is expected to reach $\sim 10\mu$as for bright
stars with $m_K \la 16.3$; while it could be $\sim 200\mu$as for stars fainter
than $m_K\sim 18.8$ \citep{Stone12}. 

The paper is organized as follows. We describe the motion equations in the Kerr
metric in Section~\ref{sec:motioneq}. We introduce the geometry and the
coordinate systems for the system of a star rotating around an MBH in
Section~\ref{sec:geo}. Our full GR method is summarized in
Section~\ref{sec:GRmethod} for calculating the relativistic motion of the star
and the photon propagation from the star to the observer.\footnote{The photon
propagation effects are important for accurately measuring the MBH spin as
pointed out by \citet{AS10b}, which are also shown in Figures~\ref{fig:f2} and
\ref{fig:f3} and discussed in Section~\ref{sec:analytical} of this paper
below.} We numerically generate mock observations for stars rotating around the
GC MBH, covering a large parameter space of stellar orbits as well as the
position and the redshift precisions.  With those mock observations, we
illustrate how the GC MBH spin and its direction can be simultaneously
constrained by using the Markov Chain Monte Carlo (MCMC) fitting technique. We
present our results on the quality of the possible spin constraints by future
facilities in Section~\ref{sec:results}.  A summary and discussion are given in
Section~\ref{sec:summary}.

\section{Motion equations in a Kerr metric} 
\label{sec:motioneq}

In the Boyer-Lindquist (BL) coordinates
$(t,r,\theta,\phi)$\,\citep{Boyer67} for the Kerr metric, the motion
of a particle (hereafter, a star or a photon) is given by 
\begin{eqnarray}
\Sigma\frac{dr}{d\tau}&=&\pm\sqrt{R},\label{eq:motionr} \\
\Sigma\frac{d\theta}{d\tau}&=& \pm\sqrt{\Theta},\label{eq:motiontheta} \\
\Sigma\frac{d\phi}{d\tau}&=& -a+\frac{\lambda}{\sin^2\theta}+\frac{aT}{\Delta},
\label{eq:motionphi} \\
\Sigma\frac{dt}{d\tau}&=&-a^2\sin^2\theta+a\lambda+\frac{(r^2+a^2)T}{\Delta},
\label{eq:motiont} 
\end{eqnarray} 
with
\begin{eqnarray}
\Sigma&=&r^2+a^2\cos^2\theta,\\
\Delta&=&r^2-2r+a^2,\\
T&=&r^2+a^2-\lambda a,\\
R&=&(1-\xi^2)r^4+2\xi^2r^3+[a^2(1-\xi^2)-q^2 \\
 & &-\lambda^2]r^2+2[(a-\lambda)^2+q^2]r-a^2q^2,\\
\Theta&=&q^2-\left[a^2(\xi^2-1)+\frac{\lambda^2}{\sin^2\theta}\right]\cos^2\theta.
\end{eqnarray} 
Here $\tau$ is an affine parameter, $a$ is the dimensionless spin parameter,
$\xi=m/E$,
$\lambda=L_z/E$, and $q^2=Q/E^2$\,\citep{Chandra83}.  The four motion constants
of the particle are its rest mass ($m$), energy at infinity ($E$), azimuthal
angular momentum ($L_z$), and Carter's constant ($Q$) (see eq.~182 in Chapter 7
in \citealt{Chandra83}).  For simplicity, the natural units are adopted above,
i.e., the gravitational constant $G$, the speed of light $c$, and the MBH mass
$\bh$ are all set to one.

\begin{figure}
\centering
\includegraphics[scale=0.6]{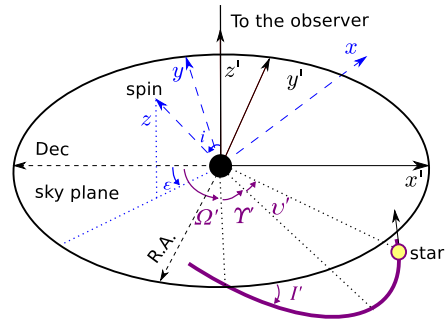}
\caption{Schematic diagram for a star rotating around the GC MBH.  A
pseudo-Cartesian coordinate system $(x',y',z')$ is defined in the distant
observer's rest frame (ORF), originating at the MBH,
with $\vec{x}'$ representing the reference direction on the sky plane $x'y'$,
and $\vec{z}'$ being the line of sight.
The directions of the declination
and the right ascension are $-\vec{x}'$ and $-\vec{y}'$.
Another pseudo-Cartesian coordinate system $(x, y, z)$ is in the MBH frame,
defined with its origin at the MBH, $\vec{z}$ representing the spin direction of
the MBH, and $\vec{y}$ being the line of intersection of the MBH equatorial
plane with the $x'y'$ plane.  The spin direction is described by two angles,
i.e., $i$
between $\vec{z}$ and $\vec{z}'$, and $\epsilon$ between the
projection vector of $\vec{x}$ on the $x'y'$ plane and
$\vec{x}'$. The stellar orbit is approximated as a Newtonian ellipse in
the ORF, determined by its six orbital elements: semimajor axis
$(a_{\rm orb})$, eccentricity $(e_{\rm orb})$, longitude of ascending node
$(\Omega')$, position angle of periapsis $(\Upsilon')$, true anomaly
$(\upsilon')$, and inclination $(I')$.  }
\label{fig:f1}
\end{figure}

\section{Geometry of the star-MBH system}
\label{sec:geo}

Figure~\ref{fig:f1} illustrates the geometry of the system composed of a star
rotating around the GC MBH with mass $\bh$ and distance of $R_{\rm GC}$ to the
sun. We define two pseudo-Cartesian coordinate systems, i.e., $(x, y, z)$ and
$(x',y',z')$. The former is in the MBH frame with $\vec{z}$ representing the
spin direction, and the latter is in the observer's rest frame (ORF). As seen
from the figure, the $(x, y, z)$ frame can be obtained by rotating the
$(x', y', z')$ frame first around the $y'$-axis clockwisely by an angle of $i$
and then around the $z'$-axis counter-clockwisely by an angle of $\epsilon$.
For a star rotating around an MBH with a semimajor axis much larger than $r_{\rm
g}\equiv G\bh/c^2$, its orbit in the ORF can be instantaneously approximated as
a Newtonian ellipse, and its initial position and velocity are determined by
its six orbital elements $(a_{\rm orb}, e_{\rm
orb}, I', \Omega', \Upsilon', \upsilon').$\footnote{The position angle of
periapsis $\Upsilon'$ is related to the argument of periapsis (denoted by
$\omega'$) by $\omega'=\pi+\Upsilon'$. The changes of the two angles are the
same, i.e., $\delta\omega'=\delta\Upsilon'$.}

The initial position of the star in the local non-rotating rest frame (LNRF)
with $(x,y,z)$ coordinates at time $t_{\star,0}$ can be converted to the BL
coordinates $(t_{\star,0}, r_{\star,0}, \theta_{\star,0}, \phi_{\star,0})$;
and the initial tetrad velocity ${\bf
u_{\star}}_{,0}=(u^t_{\star,0},,u^r_{\star,0}, u^\theta_{\star,0}, u^\phi_{\star,0})$
of the star in the BL coordinates can be obtained from the three-velocity ${\bf
v}_{\star,0}=(v^r_{\star,0}, v^{\theta}_{\star,0}, v^{\phi}_{\star,0})$ in
the LNRF\,\citep[see][]{Chandra83}. The motion constants $\lambda$, $q$, and
$\xi$ can also be obtained from the position and the tetrad velocity of the
star.

\section{Full General Relativistic Numerical Method}
\label{sec:GRmethod}

We use the full GR numerical method developed in \citet{ZLY15} to calculate
both the stellar motion around the GC MBH and the photon propagation from the
star to the observer.

For a star rotating around an MBH with any set of $(\xi,\lambda,q^2)$, its
orbital evolution can be obtained by numerically solving
Equations~(\ref{eq:motionr})-(\ref{eq:motiont}).  We adopt the explicit
fifth (fourth)-order Runge-Kutta method \citep{Hairer93} to integrate these
equations, and we set high relative accuracies $(\leq10^{-12})$ for those
integral quantities in order to calculate the spin-induced effects precisely.

For a photon from the star received by a distant observer located at
$(r, \theta, \phi)=(R_{\rm GC}, i, 0)$, its trajectory may be bended and its energy
at the receiving time may be shifted away from that at the emission time.  The
position of the star on the observer's sky plane can be described by impact
parameters $\alpha$ and $\beta$, with the former representing the displacement
of the star image perpendicular to the projected symmetry axis of the MBH and
the latter representing the displacement parallel to this axis. The motion
constants are $\xi=0$, $\lambda=-\alpha\sin i$, and $q^2=\beta^2+(\alpha^2
-a^2)\cos^2 i$ for a photon received by the observer at
$(\alpha,\,\beta)$\,\citep{Chandra83}.  We adopt the ray-tracing technique
(\citealt{Rauch94}, \citealt{LY01}, for details see \citealt{ZLY15}) to search
the parameter space $(\alpha,\,\beta)$ for those photons received by the
observer and emitted from a star on a relativistic orbit.  As a result, the
position of the star on the sky plane $(\alpha_k,\,\beta_k)$ at a given
observing time $t_{{\rm obs},k}$ is connected to the position of the star at
$(t_k, r_k, \theta_k,\,\phi_k)$. In the mean time, the tetrad-momentum of the
photon ${\bf p}_k$ can also be obtained, and the redshift (measured through the
shift of emission or absorption lines) $Z_k=-{{\bf p}_k\cdot{{\bf
u}}_{\star,k}}/E_0-1$, where $E_0$ is the energy of the photon in the rest
frame of the star at the emission time, ${\bf u}_{\star,k}$ is the tetrad
velocity of the target star at time $t_{{\rm obs},k}$. Therefore, the
observational quantities of the star (position trajectory and redshift curve)
can be mapped to its relativistic orbital motion. 

For the full GR numerical calculations below, we set the numerical accuracies
for  positions and redshifts (or velocities) to be $10^{-4}\mu$as and
$10^{-6}\kms$, respectively.

\begin{figure*} 
\includegraphics[scale=0.55]{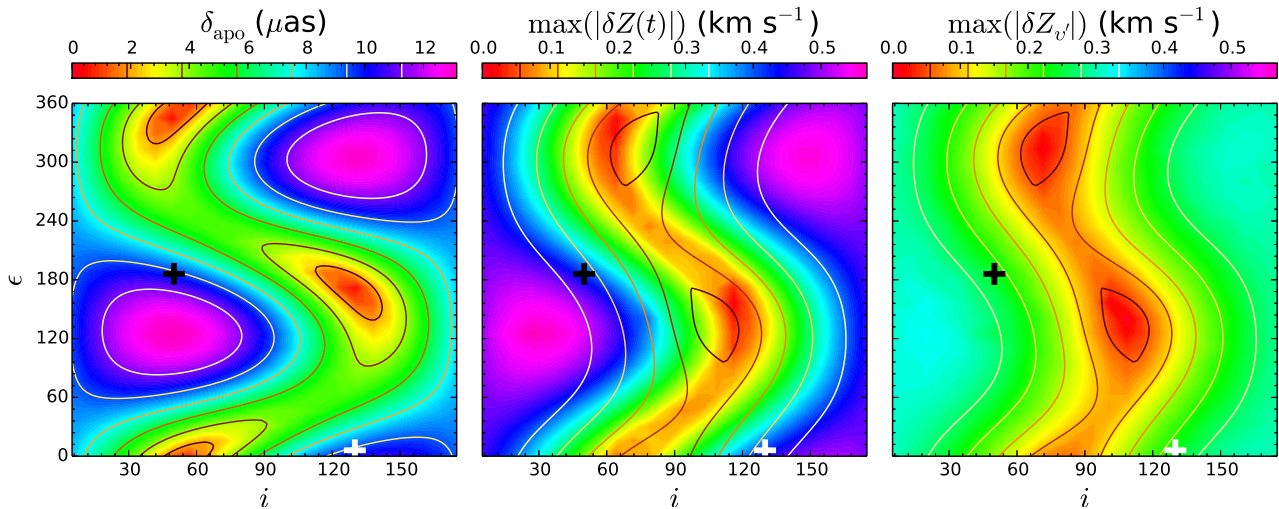}
\centering

\caption{Dependence of the spin-induced signals on the spin direction
$(i,\epsilon)$ for S2/S0-2 (see Fig.~\ref{fig:f1}) by assuming the MBH spin
$a=0.99$. The left and the middle panels show the spin-induced position
displacements at apoapsis and the maximum values of the spin-induced redshift
differences (defined as $\delta Z(t)$ in Equation~\ref{eq:deltaz}) after the
motion of one full orbit, respectively. The orbital true anomalies $\upsilon'$
at the same time $t$ are slightly different for different spins and the
spin-induced signals caused by the difference in true anomalies are included in
the middle panel. For comparison with some analytical results, the right panel
shows the maximum redshift differences caused by the spin-induced changes of
the orbital plane (defined as $\max(|\delta Z_{\upsilon'}(t)|)$, see
Eq.~\ref{eq:redshift}), with the signals caused by the difference in true
anomalies removed.  In each panel, the color maps represent the results
obtained from the full GR calculations with color indices shown at the top, and
the color contour lines represent the analytical ones estimated from
Equation~(\ref{eq:position}) in the left panel and Equation~(\ref{eq:redshift})
in the middle and the right panels.  The ``{\bf +}'' symbols mark the normal of
the orbital plane of the clockwise rotating stellar disk located at the
sub-parsec scale of the GC \citep{Yelda2014}. As seen from the figure, the
strength of the maximum and the minimum of the spin-induced signals can differ
by up to two orders of magnitude, and the spin-induced signals are close to the
maximum signals if the spin direction is the same as the normal of the young
stellar disk.  See Sections~\ref{sec:numerical} and \ref{sec:analytical}.  }
\label{fig:f2}
\end{figure*}

\begin{figure*}
\includegraphics[scale=0.55]{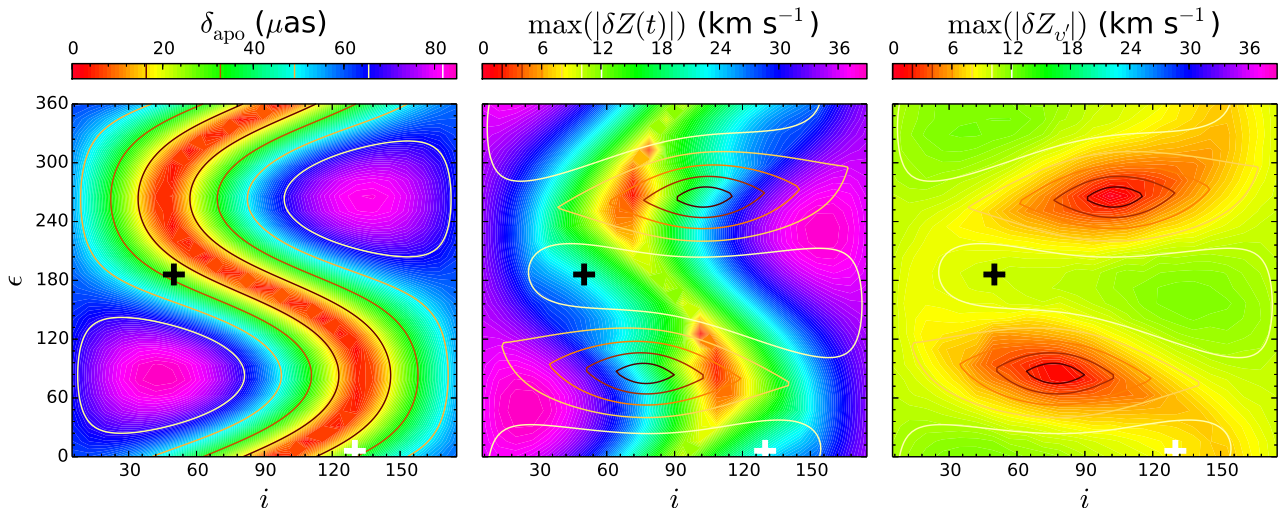}
\centering
\caption{Legends similar to Fig.~\ref{fig:f2}, but for a star with a semimajor
axis of $a_{\rm orb}=300\AU$, an eccentricity of $e_{\rm orb}=0.95$, and the
same orbital orientation as that of S0-102.  As seen from the
figure, the strength of the spin-induced signals are at the same order of
magnitude as the maximum strength if the spin direction is the same as the
normal of the young stellar disk. The spin-induced signals are stronger than
those shown in Figure~\ref{fig:f2}.}
\label{fig:f3}
\end{figure*}

\section{Results}
\label{sec:results}

\subsection{Numerical results from full GR calculations}
\label{sec:numerical}

We show that the spin-induced effects on the motion of a given star are
sensitive to the spin direction in Figures~\ref{fig:f2} and \ref{fig:f3},
where the spin-induced signals are obtained by the difference of the signals
between the cases of the MBH spin $a=0.99$ and $a=0$.

Figure~\ref{fig:f2} shows the dependence of the spin-induced signals on the
spin directions for S2/S0-2 after the motion of one full orbit, where the
initial orbital elements of the star  are set to $a_{\rm orb}=984\AU$, $e_{\rm
orb}=0.88$, $I'=135\arcdeg$, $\Omega'=225\arcdeg$, and $\Upsilon'=63\arcdeg$
\citep{Gillessenetal09}.  The left panel of Figure~\ref{fig:f2} shows the
position displacements at the apoapsis of S2/S0-2 between the case with a
rapidly rotating MBH ($a=0.99$) and that with a non-rotating MBH; and the
middle panel shows the maximum value of the spin-induced redshift differences
over one full orbital period between these two cases. As seen from the left
panel, the spin-induced position displacement at apoapsis is most significant,
i.e., $\sim13\microas$ for $a=0.99$ when
$(i,\,\epsilon)=(49\arcdeg,\,125\arcdeg)$ or $(131\arcdeg,\,305\arcdeg)$, and
the least significant, i.e., $\sim 0.2\microas$ for $a=0.99$ when
$(i,\,\epsilon) = (50\arcdeg,\,351\arcdeg)$ or $(130\arcdeg,\,171\arcdeg)$. As
seen from the middle panel, the maximum value of the spin-induced redshift
differences over a full orbit is the most significant, i.e., $0.56 \kms$ for
$a=0.99$ when $(i,\epsilon)=(28\arcdeg,\,127\arcdeg)$ or
$(152\arcdeg,\,307\arcdeg)$, and the least significant, i.e., $\sim
10^{-2}\kms$ if $(i,\epsilon)=(118\arcdeg,144\arcdeg)$ or
$(72\arcdeg,324\arcdeg)$. As seen from the figure, the normal of the clockwise
rotating stellar disk is close to the direction of the spin with the maximum
spin-induced signals.  If the spin direction is the same as the normal of the
clockwise rotating stellar disk, the spin-induced apparent position
displacement at the apoapsis and the maximum redshift difference are close to
the above maximum values (see Fig.~\ref{fig:f2}); and accurate measurements of
the motion of the star are helpful to test whether the two directions are the
same or significantly different.

One may note here that constraining the spin direction is also important as it
may record the assembly history of the MBH. For example, if the spin direction
is the same as the normal of the young stellar disk in the GC, it may indicate
that the growth of the GC MBH via the previously existing gas disk, which
coincides with the stellar disk, is important; or if the spin direction is
significantly different from the normal of the stellar disk, which might
suggest that the accretion episodes that happened at a later time are chaotic
and the growth of the MBH via a single accretion episode is not significant.

Similar to Figure~\ref{fig:f2}, Figure~\ref{fig:f3} shows the dependence of
spin-induced signals on the spin directions for a hypothetical star after the
motion of one full orbit by assuming the initial orbital elements $a_{\rm
orb}=300$\,AU, $e_{\rm orb}=0.95$, $I'=151\arcdeg$, $\Omega'=175\arcdeg$, and
$\Upsilon'=185\arcdeg$. The orbital orientation of this star is set to be the
same as that of S0-102 \citep{Meyer12}. 
The orbital period of such a star is $\sim 2.6\yr$.
As seen from the left panel of the figure, the
spin-induced position displacement at apoapsis is most significant, i.e., $\sim
84.9\microas$ for $a=0.99$ when $(i,\,\epsilon)= (42\arcdeg,\,78\arcdeg)$ or
$(138\arcdeg,\,258\arcdeg)$, and the least significant, i.e., $\sim
1.7\microas$ for $a=0.99$ when $(i,\,\epsilon)=(116\arcdeg,\,140\arcdeg)$ or
$(64\arcdeg,\,320\arcdeg)$. As seen from the middle panel, the maximum value of
the spin-induced redshift differences over a full orbit is the most
significant, i.e., $38.9\kms$ for $a=0.99$ when
$(i,\epsilon)=(27\arcdeg,\,47\arcdeg)$ or $(153\arcdeg,\,227\arcdeg)$, and the
least significant, i.e., $\sim 1.3\kms$ if
$(i,\epsilon)=(101\arcdeg,125\arcdeg)$ or $(79\arcdeg,305\arcdeg)$.  If the
spin direction is the same as the normal of the clockwise rotating stellar
disk, the spin-induced apparent position displacement at the apoapsis is about
one third of the above maximum value, and the spin-induced redshift difference
is about half of the above maximum value.

\begin{figure}
\centering \subfigure[]{\includegraphics[scale=0.40]{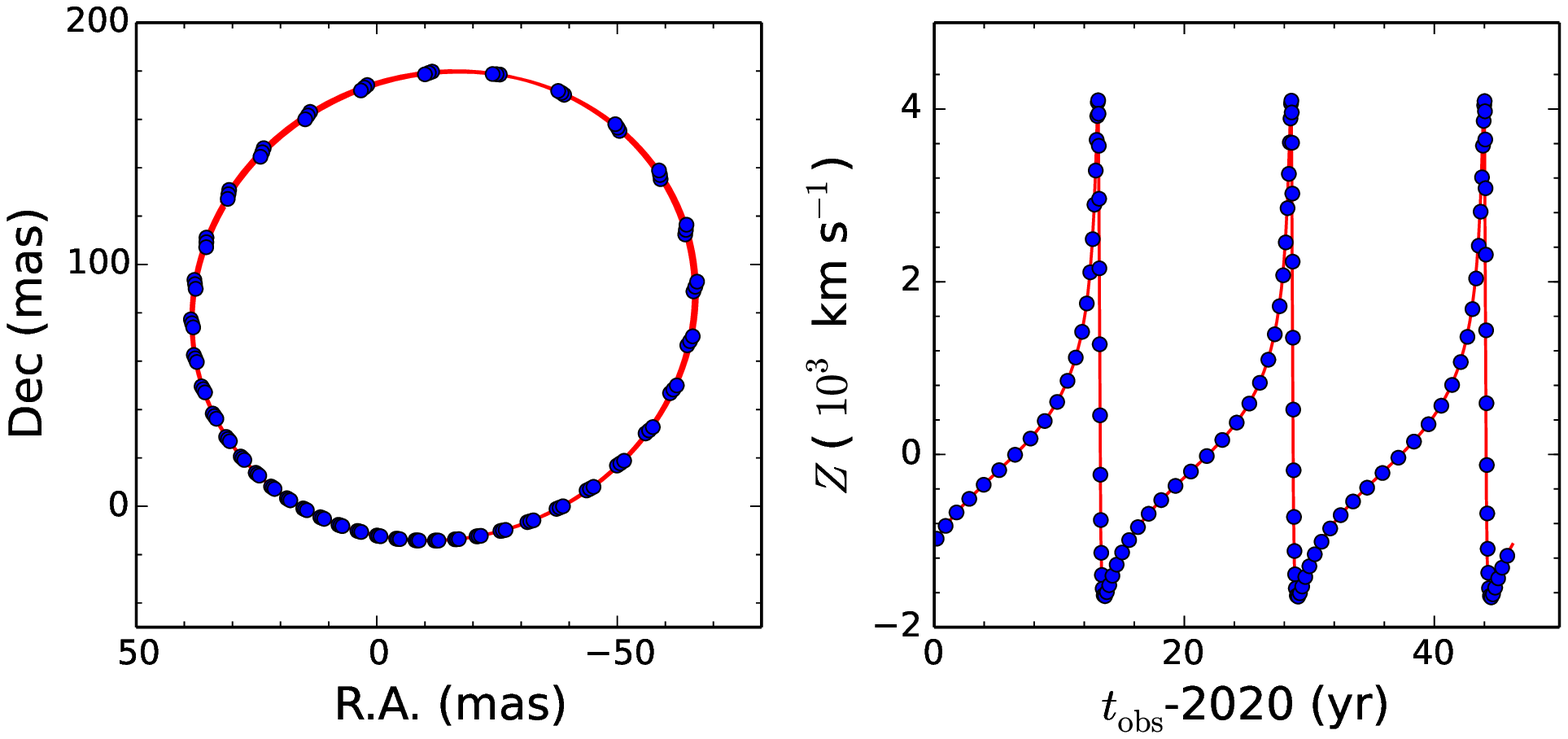}}
\subfigure[]{\includegraphics[scale=0.40]{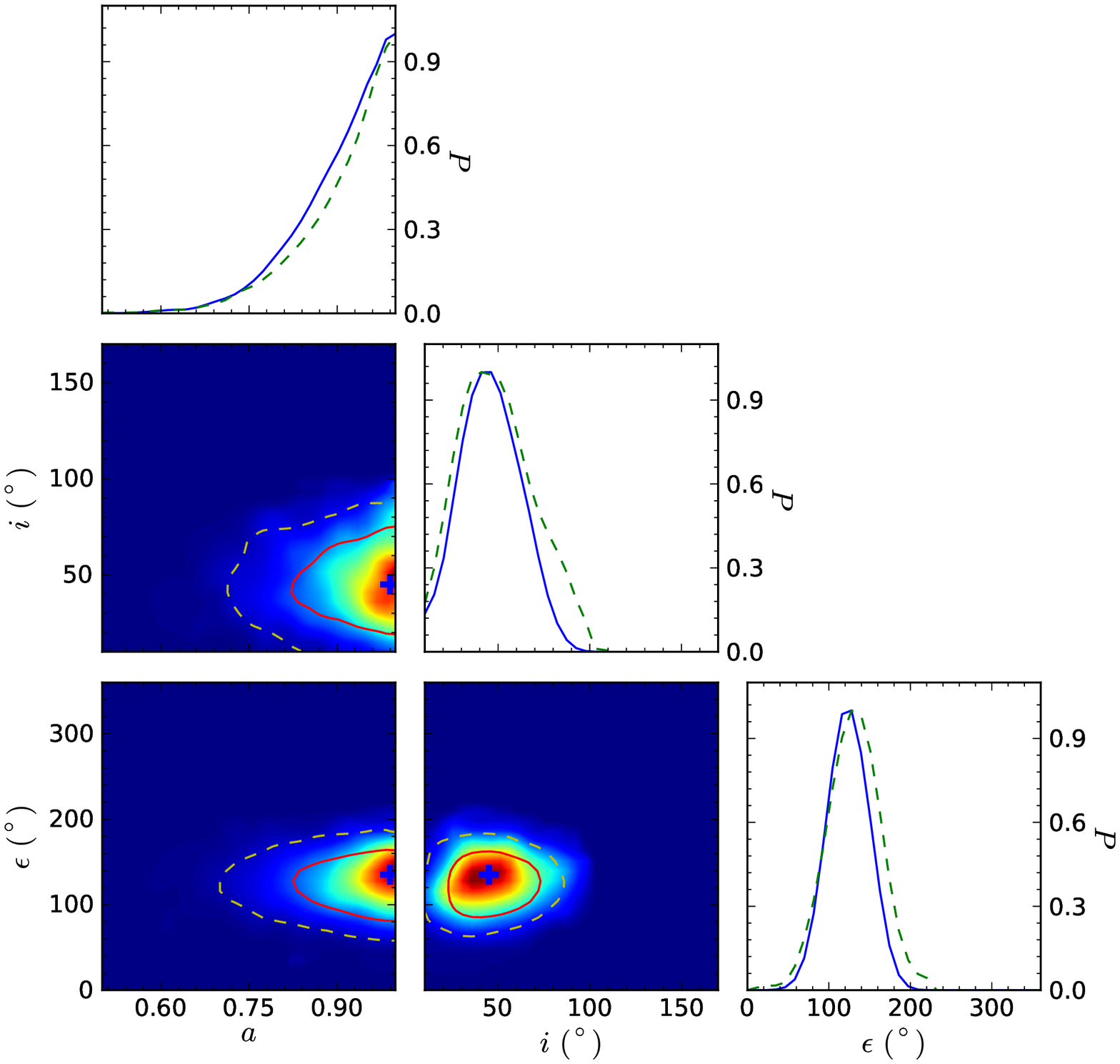}} 
\caption{Best fits to the mock observations of S2/S0-2.  Panel (a) shows
the best fits (red lines) to the mock observations (blue solid circles) on the
apparent positions (left) and redshifts (right).  Panel (b) shows the 2D
contours and the 1D probabilities of parameters $(a,i,\epsilon)$ obtained from
the fitting of the mock observations of S2/S0-2 over three full orbits by
assuming intrinsically $(a,i,\epsilon)= (0.99,49\arcdeg,125\arcdeg)$.  In the
2D maps, the red solid and the yellow dashed lines show the
$1$-$\sigma$ and the $2$-$\sigma$ confidence levels, respectively; in the
panels for the 1D probabilities, the blue solid and the green dashed lines
show the 1D marginalized distribution and the 1D mean likelihood,
respectively.  This figure illustrates an example that the spin of the GC MBH
and its direction can be constrained well. See details in
Section~\ref{sec:example}. }
\label{fig:f4}
\end{figure}

\begin{figure*} 
\centering
\includegraphics[scale=0.70]{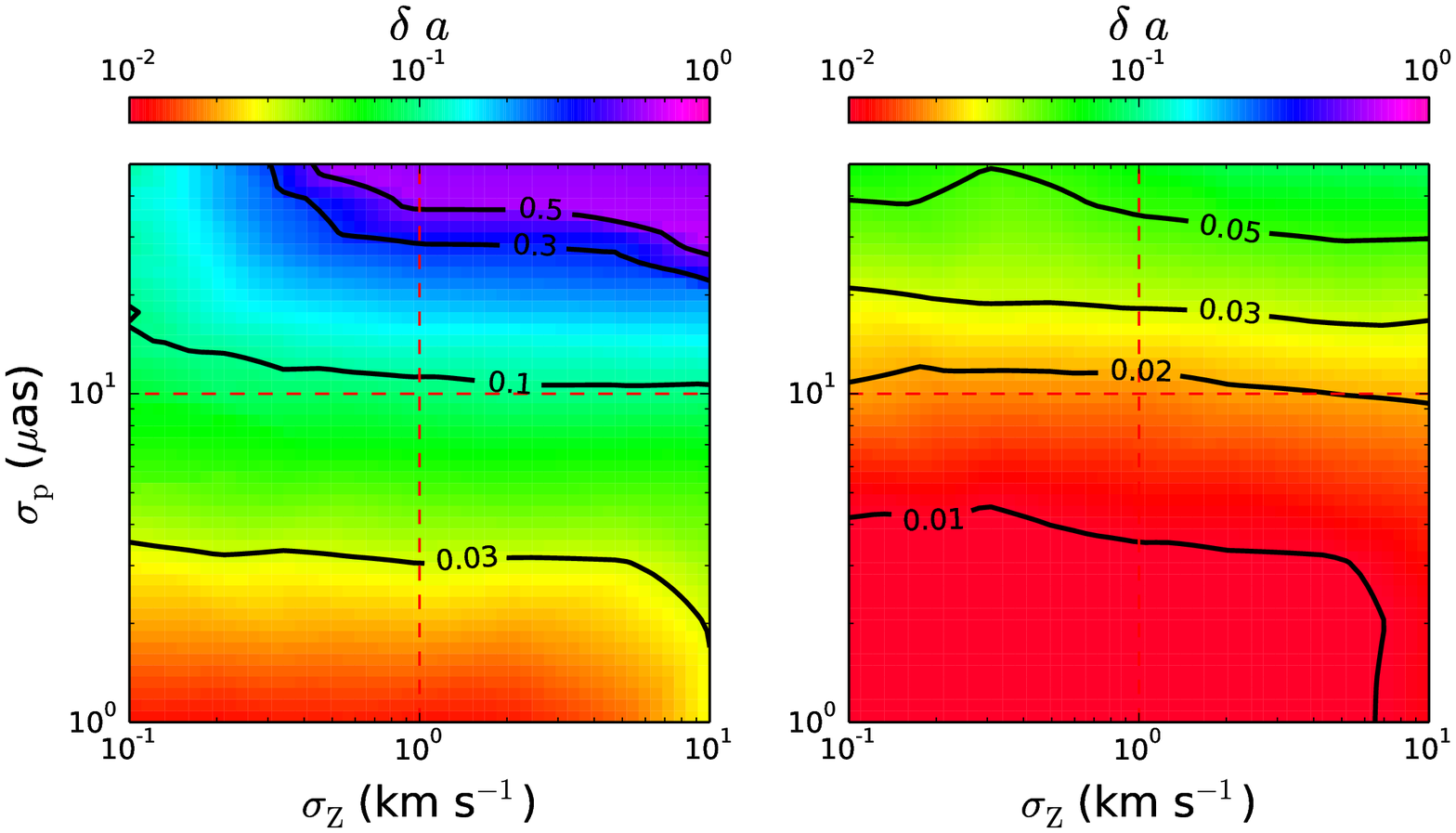}
\caption{Dependence of the quality of the MBH spin constraints ($\delta a$)
on the astrometric precision ($\sigma_{\rm p}$) and the velocity precision
($\sigma_z$), which are obtained from mock observations of two different stars
over three full orbits. The left panel shows the result obtained from the mock
observations of S2/S0-2 by
assuming $(a,\,i,\,\epsilon)=(0.99,\,49\arcdeg,\ 125\arcdeg)$; and the right
panel shows the result obtained from the mock observations of 
a hypothetical star with $(a_{\rm orb},e_{\rm orb})=(300\AU, 0.95)$ and an
orientation that is the same as that of S0-102, by assuming
$(a,\,i,\,\epsilon)=(0.99,\,49\arcdeg,\ 125\arcdeg)$. The top color bars 
indicate the value of $\delta a$. The black lines represent
the contours of $\delta a$ with the marked values. The red dashed
lines in both panels mark the astrometric and the velocity precisions to
be possibly achieved by the next-generation facilities. As seen from the
figure, in the ranges of astrometric precision and velocity
precision with 1--30 $\mu$as and 0.1--10$\kms$, an improvement in astrometric
precision would be more effective at improving the quality of constraining the
spin than an improvement in velocity precision. 
}
\label{fig:f5}
\end{figure*} 

\subsection{Comparison with analytical approximations}
\label{sec:analytical}

\subsubsection{Analytical approximations}

{\it Orbital precession:} the leading spin effect on the motion of a star is
the Lense-Thirring (LT) precession of the star's orbital plane, which causes
the changes of the longitude of ascending node ($\Omega$) and the position
angle of periapsis ($\Upsilon=\omega-\pi$) per orbit by 
\begin{equation}
\delta\Omega_{\rm
LT}=4\pi a a^{-3/2}_{\rm orb} (1- e^2_{\rm orb})^{-3/2}
\end{equation}
and 
\be
\delta{\Upsilon}_{\rm LT}=-3 \delta \Omega_{\rm LT}\cos\zeta,
\ee
respectively\,\citep[e.g.,][]{LT18,Mashhoon84,Wex99}. Here $\Omega$ and $\Upsilon$ are defined
for the star relative to the equatorial plane of the Kerr BH in the $(x,y,z)$
frame, 
\be
\zeta=\arccos(\hat{\bf a}\cdot\hat{\bf n}_*)=\arccos[\cos I'\cos i -
\sin I'\sin i\sin(\Omega'-\epsilon)],
\ee
where $\hat{\bf a}$ is the unit vector along
the spin direction, $\hat{\bf n}_*$ is the normal vector of the star's orbital
plane, and $a_{\rm orb}$ is in units of $r_{\rm g}$. The quadrupole (Q) term of
the MBH can also lead to orbital precession by 
\be
\delta\Omega_{\rm Q} = -3\pi
a^2 \cos \zeta a^{-2}_{\rm orb} (1-e^2_{\rm orb})^{-2}
\ee
 and
\be
\delta\Upsilon_{\rm Q} = 1.5\pi a^2 (1-5\cos^2\zeta) a^{-2}_{\rm orb}
(1-e^2_{\rm orb})^{-2}
\ee
 \citep[e.g.,][]{Barker75, Wex99}. The changes of
$\Omega$ and $\Upsilon$ are then approximately given by 
\be
\delta\Omega =
\delta\Omega_{\rm LT} + \delta\Omega_{\rm Q}
\ee
 and 
\be
\delta\Upsilon =
\delta\Upsilon_{\rm LT} + \delta\Upsilon_{\rm Q}.
\ee
  Considering the projection
effects, the changes of $\Omega'$, $\Upsilon'$, and $I'$ are
\be
\delta\Omega'=\left[\cos i + \cos I'\sin i\sin(\Omega'-\epsilon)/\sin
I'\right]\delta\Omega,
\ee
\be
\delta\Upsilon'= - \sin i\sin(\Omega'-\epsilon)/\sin
I' \delta\Omega + \delta \Upsilon,
\ee
 and 
\be
\delta I'= - \sin
i\cos(\Omega'-\epsilon)\delta\Omega,
\ee
 respectively.

{\it Position displacement at apoapsis:} after the motion of one full orbit,
the position displacement at the apoapsis of a star rotating around an MBH with
spin $a$ is approximately
\begin{eqnarray} 
\delta_{\rm apo}&\sim& a_{\rm orb}(1+e_{\rm orb})[
(1-\sin^2\Upsilon'\sin^2I')\delta^2\Omega'  \nonumber \\ & &  +(1-
\cos^2\Upsilon'\sin^2 I')\delta^2\Upsilon'  \nonumber \\ & & 
+2\cos I'\delta\Omega'\delta \Upsilon'+\sin^2\Upsilon'\sin^2I'\delta^2I'
 \nonumber \\ & & -\frac{1}{2}\sin2\Upsilon'\sin
2I'\delta\Upsilon'\delta I'-\sin 2\Upsilon'\sin I'\delta\Omega'\delta
I']^{1/2}.  \nonumber \\
\label{eq:position} 
\end{eqnarray} 
When the quadrupole term is not significant, we have $\delta_{\rm apo} \propto
a_{\rm orb}^{-1/2} (1-e_{\rm orb}^2)^{-3/2}$.

{\it Redshift difference:} we consider two cases, one for a star with given initial orbital
elements rotating around a non-spinning MBH with mass ($M_\bullet$),
and the other for a star with the same initial orbital elements but
rotating around a rapidly rotating MBH with the same mass
($M_\bullet$). The redshift difference between these two cases at a
given time $t$ measured by a distant observer is given by
\be
\delta Z(t) \sim \delta Z_{\upsilon'}|_{\rm LT+Q} + \delta
Z_{\delta \upsilon'},
\label{eq:deltaz}
\ee
where $\delta Z_{\upsilon'}|_{\rm LT+Q}$ is the redshift
difference at a fixed $\upsilon'$ due to the spin-induced (LT and Q) precessions
and $\delta Z_{\delta \upsilon'}\equiv \frac{\partial
Z_{\upsilon'}}{\partial \upsilon'} \delta \upsilon' $ is the redshift
difference due to the slight difference between the true anomalies
($\upsilon'$) of these two stars at a given time $t$ caused by the
difference of the spin values. The two terms in
Equation~(\ref{eq:deltaz}) can be roughly given by the projection of
the spin-induced Keplerian velocity difference at the line of sight,
i.e.,
\begin{eqnarray}
\delta Z_{\upsilon'}|_{\rm LT+Q}  & \sim &
-\left[\frac{G\bh}{a_{\rm orb}(1-e^2_{\rm orb})}\right]^{1/2}
\left\{\cos I' [e_{\rm orb}\cos\Upsilon' \right. \nonumber \\
& & \left.  +\cos(\Upsilon'+\upsilon')]\delta I' -\sin I'[e_{\rm
orb}\sin\Upsilon' \right.\nonumber \\ 
& & \left.  +\sin (\Upsilon'+\upsilon')]\delta\Upsilon'\right\},
\label{eq:redshift}
\end{eqnarray} 
\be
\delta Z_{\delta \upsilon'} \sim -\left[\frac{G\bh}{a_{\rm
orb}(1-e^2_{\rm orb})}\right]^{1/2} \sin I'\sin(\Upsilon'+\upsilon')
\delta \upsilon'(t).
\label{eq:deltazv}
\ee
The maximum difference of $\delta Z_{\upsilon'}|_{\rm LT+Q}$ over a
full orbit can be obtained from Equation~(\ref{eq:redshift})
analytically (see \citealt{ZLY15}). However, $\delta \upsilon'$ may
only be obtained numerically for a star with arbitrary orbital
elements. For the following comparisons between the numerical and
the analytical results of the spin-induced redshift differences, we
therefore mainly focus on $\delta Z_{\upsilon'}|_{\rm LT+Q}$.
Note that $\delta Z_{\upsilon'}|_{\rm LT+Q} \propto a^{-2}_{\rm orb} (1-e_{\rm
orb}^2)^{-2}$ if $e_{\rm orb}\rightarrow 1$, and $\delta Z_{\upsilon'}|_{\rm
LT+Q} \propto a^{-2}_{\rm orb}$ if $e_{\rm orb} \rightarrow 0$, 
which are consistent with the findings by \citet{AS10b}. 

In the derivation of Equations~(\ref{eq:position}) and (\ref{eq:redshift}), we
omit the effects due to photon propagation from the star to the distant
observer.  For those photons emitted from the star at different locations,
their trajectories may be bended and their energy may be shifted differently.

\subsubsection{Comparison}

As shown in the left panels of Figures~\ref{fig:f2} and \ref{fig:f3}, the
position displacements at the apoapsis obtained from
Equation~(\ref{eq:position}) for S2/S0-2 and the hypothetical star are quite
consistent with the numerical ones. The residuals between the full GR results
and the analytical ones are on the order of one percent (for S2/S0-2) to ten
percent (for the hypothetical star) of the full GR results, which are mainly
caused by the omission of the photon propagation effects and the high-order
precessions.

The middle panels of Figures~\ref{fig:f2} and \ref{fig:f3} show both the full
GR results on the spin-induced maximum redshift differences over one full orbit
and the analytical ones due to the LT and Q precessions (Eq.~\ref{eq:redshift})
for S2/S0-2 and the hypothetical star, respectively. Apparently, the numerical
results deviate significantly from the analytical ones derived from
Equation~(\ref{eq:redshift}), mainly because of the omission of the second term
in Equation~(\ref{eq:deltaz}) in the analytical estimates. The right panels of
Figures~\ref{fig:f2} and \ref{fig:f3} show the full GR results on the
spin-induced redshift difference obtained by numerically removing the part
$\delta Z_{\delta \upsilon'}$ for S2/S0-2 and the hypothetical star,
respectively. As seen from these two panels, the full GR results are quite
consistent with the analytical ones.  However, our calculations show that the
residuals between the full GR results and the analytical results are on the
order of a few to a few tens percent for both S2/S0-2 and the hypothetical
star, mainly due to the omission of the photon propagation effects and the
high-order precessions.  Our calculations show that the spin-induced effects on the photon propagation and
the rotation velocity at a given position can introduce some differences in the
redshifts, though about $5-10$ times smaller than that introduced by the
position shift (Eq.~\ref{eq:redshift}) for S2/S0-2 and the hypothetical star.
Therefore, the photon propagation effects cannot be neglected when considering
an accurate constraint on the MBH spin and metric.

The contribution of the Q precession (e.g., $\delta\Omega_{\rm Q}$ and
$\delta\Upsilon_{\rm Q}$) to the spin-induced effects is only about one to a
few percent, substantially smaller than that contributed by the LT precession,
for stars with semimajor axes $\sim 300-1000\AU$ and eccentricities $\ga 0.88$.
Our calculations show that the deviations of the analytical results on the
spin-induced position displacements and redshift differences from those
numerical ones are on percentage or even a higher level, comparable to the
contribution from the quadrupole moment, mainly because of the omission of the
photon propagation effects and the high-order precessions in obtaining the
analytical results (see details in \citealt{ZLY15} and \citealt{A10a}).
Therefore, it is important to use the full GR numerical method to account for
the photon propagation effects when considering constraints on the MBH spin
with high precision (e.g., percentage level) or the quadrupole moment.

\subsection{Orbital reconstruction and the quality of constraining the
MBH spin}
\label{sec:mcmc}

Using the full GR method introduced above, we can generate mock observations on
both the trajectory of the apparent position and the redshift curve measured by
a distant observer for any star, e.g., S2/S0-2 or S0-102, with any given set of
astrometric and velocity precisions $(\sigma_{\rm p},\,\sigma_Z)$; and then we
use the mock observation results to reconstruct orbital elements of the star
and the intrinsic properties of the MBH by the MCMC fitting technique as
follows. The trajectory and the redshift curve of a star are determined by 11
parameters (${\bf\Theta}$), including the MBH properties
$(M_\bullet,a,i,\epsilon)$, the distance from the MBH to the sun $(R_{\rm
GC})$, and the star's initial orbital elements $(a_{\rm orb,0},e_{\rm
orb,0},\Omega'_0,I'_0,\Upsilon'_0, \upsilon'_0)$.\footnote{Note that here we
ignore the possible velocity of the MBH relative to Sgr A* and the acceleration
of the MBH with respect to the observer. In principle, both the velocity
and the acceleration can be added to the fitting as free parameters and
constrained by the observations. By doing so, the motion of the solar system
may also be constrained simultaneously with a high accuracy, since the pattern
of the relative motion of the observer to the GC MBH is different from the
spin-induced motion. See a detailed discussion in
\citet{ZLY15}. } 

To obtain mock observations, we set
$M_\bullet=4\times10^6\msun$, $R_{\rm GC}=8{\rm\,kpc}$. For a given set of mock
observations on both apparent position and redshift ${\bf D}=(\alpha_{{\rm
obs},k},\beta_{{\rm obs},k},Z_{{\rm obs},k})$ at time $t_{{\rm obs},k}$
($k=1,2, ...,N$ and $N$ is the total number of the mock observations), we can
use the full GR method to generate model results on $(\alpha_k,\beta_k,Z_k)$ at
$t_{{\rm obs},k}$ for any given set of parameters ${\bf \Theta}$. We then adopt
the MCMC technique to fit the mock observations and obtain the best fit of each
parameter, i.e., the posteri probability of the model parameters 
\be
P({\bf
\Theta}|{\bf D})\propto \exp\left( -\chi^2/2\right)P({\bf \Theta}).
\ee
Here
\begin{eqnarray}
\chi^2&=&\sum_{k_1=1}^{N_1}\frac{(\alpha_{k_1}-\alpha_{{\rm obs},k_1})^2+
(\beta_{k_1}-\beta_{{\rm obs},k_1})^2}{\sigma^2_{\rm p}} \\
& & +\sum_{k_2=1}^{N_2}
\frac{(Z_{k_2}-Z_{{\rm obs},k_2})^2}{\sigma^2_Z},
\end{eqnarray}
and $P({\bf \Theta})$ is the prior
probability distribution and assumed to be a flat distribution, $N_1$ and $N_2$
are the number of mock observations for the position and the redshift of the
target star, respectively.

\subsubsection{Examples}\label{sec:example}

We first generate mock observations for S2/S0-2 over three full orbits by
assuming $(a, i, \epsilon)= (0.99, 49\arcdeg, 125\arcdeg)$ and $(\sigma_{\rm
p}, \sigma_Z)=(10\microas, 1\kms)$. With this assumed spin direction, the
spin-induced position displacement for S2/S0-2 are the most significant.  The
total number of mock observations are set to $N_1=N_2=120$.  We take more mock
observations near the periapsis than that near the apoapsis.
Figure~\ref{fig:f4} shows both the fittings to the mock observations (panel
(a)) and the two-dimensional (2D) probability contours and one-dimensional (1D)
marginalized probability distributions for the spin parameters $(a, i, \epsilon)$
(panel (b)).  According to the fitting, we find that the spin value $a$ is
constrained to $0.90^{+0.09}_{-0.15}$, well consistent with the input one; and
the spin direction can be simultaneously constrained with
$(i, \epsilon)=(46\arcdeg^{+26\arcdeg}_{-24\arcdeg},
124\arcdeg^{+38\arcdeg}_{-38\arcdeg})$.  In the mean time, the MBH mass and the
distance from the MBH to the sun can be constrained to a relative accuracy of
$\sim 1-2\times10^{-4}$.

\begin{figure*}
\centering
\includegraphics[scale=0.70]{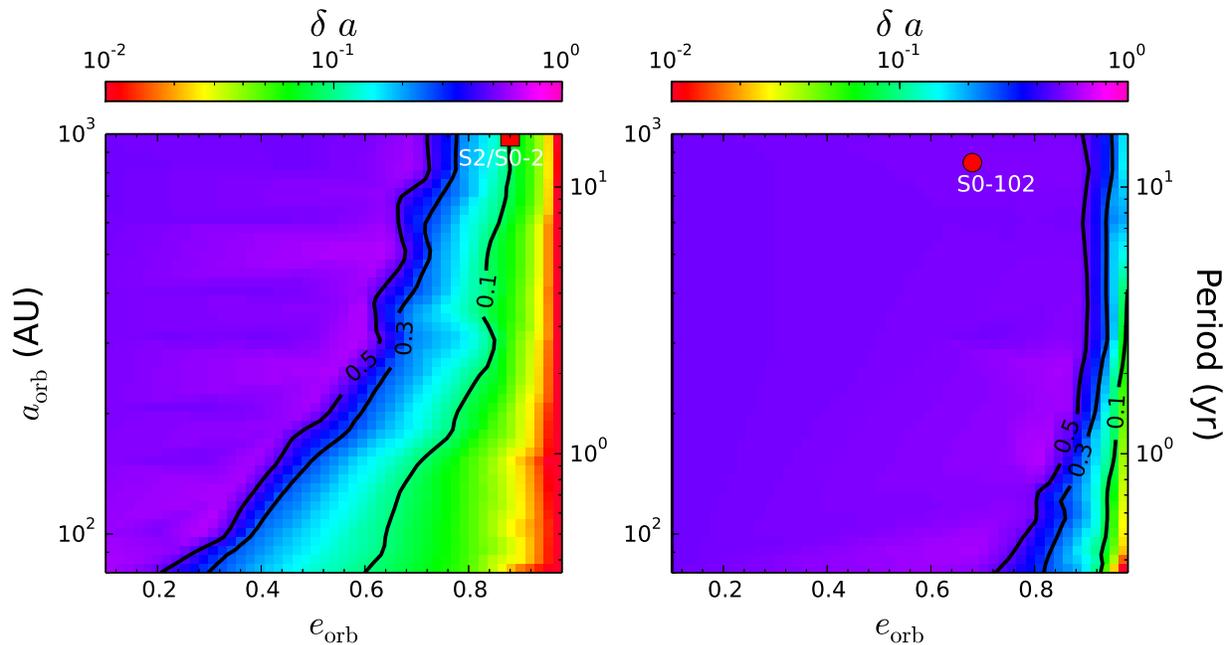}
\caption{Dependence of the quality of the spin constraints ($\delta a$) on the
semimajor axis (or the orbital period) and the eccentricity of hypothetical
stars. The left panel shows the simulation results obtained for stars with
orbital orientations that are the same as that of S2/S0-2 and assuming
$(a, i, \epsilon) = (0.99, 49\arcdeg, 125\arcdeg)$, $(\sigma_{\rm
p}, \sigma_z) = (10\microas, 1\kms)$; and the right panel shows the simulation
results obtained for those stars with orbital orientations that are the same as that of
S0-102 and assuming $(a, i, \epsilon) = (0.99, 49\arcdeg, 125\arcdeg)$,
$(\sigma_{\rm p}, \sigma_z) = (50\microas, 5\kms)$. The top color bar indicates
the value of $\delta a$. The black lines represent the contours of $\delta a$
with the marked values. The solid square in the left panel and the solid
circle in the right panel mark the positions of S2/S0-2 and S0-102,
respectively \citep{Meyer12}.  In the figure, the spin of the GC MBH can be
constrained with 1-$\sigma$ error $\la 0.1$ (right panel) or even $\la 0.02$
(left panel) by monitoring the orbital motion of a star with semimajor axis
$\la 300\AU$ and eccentricity $\ga 0.95$ over a period of $\la 10$~yr.
}
\label{fig:f6}
\end{figure*}

The spin-induced effects are moderate (see Fig.~\ref{fig:f2}) if the spin
direction is close to the orientation of the clockwise rotating stellar disk
$(i, \epsilon)=(130\arcdeg, 6\arcdeg)$ or $(50\arcdeg, 186\arcdeg)$ in the GC
\citep{Yelda2014}. For this case, the spin-induced effects on both the position
and the redshift are moderate so that the spin and its direction can still be
well constrained $(a, i, \epsilon)=
(0.70^{+0.30}_{-0.34}, 137\arcdeg^{+33\arcdeg}_{-46\arcdeg}, 
-16\arcdeg^{+97\arcdeg}_{-107\arcdeg})$ if $(\sigma_{\rm p} \sigma_Z)=(10\microas, 
1\kms)$.

If the spin direction is $(i, \epsilon)=(50\arcdeg, 351\arcdeg)$ or
$(130\arcdeg, 171\arcdeg)$, the spin-induced effects on the position
displacements are the least and those on the redshift differences are also
close to the least, and thus the spin and its direction cannot be well
constrained even if $(\sigma_{\rm p}, \sigma_Z)=(10\microas, 1\kms)$.  

By surveying the parameter space, we find that the spin can be well constrained
for about half of the possible spin directions shown in Figure~\ref{fig:f2}
by using the motion of three full orbits of S2/S0-2 if $a$ is close to 1 and
$(\sigma_{\rm p}, \sigma_Z)\sim(10\microas, 1\kms)$. With such expected
position and redshift precisions, the GC MBH spin may be better constrained via
the motion of S2/S0-2 if choosing a substantially longer monitoring period (but
impractical); however, a robust constraint on the spin would be difficult to
obtain via the motion of S2/S0-2 if choosing a substantially shorter monitoring
period (e.g., a decade) and considering that the spin is likely not to point
just at the direction with the most significant spin-induced effects.

\subsubsection{Dependence on $(\sigma_{\rm p},\sigma_Z)$}

We further generate mock observations for S2/S0-2 over three full orbits
by assuming various combinations of $\sigma_{\rm p}$ and $\sigma_Z$, and then
we adopt the MCMC technique to obtain the constraint on the MBH spin
for each set of $(\sigma_{\rm p}, \sigma_Z)$ as $a^{+\delta
a_1}_{-\delta a_2}$, where $\delta a_1$ and $\delta a_2$ are the
$1$-$\sigma$ errors obtained from each fit.  We set $\delta
a=\max(|\delta a_1|, |\delta a_2|)$ to represent the quality of the spin
constraint.  The left panel of Figure~\ref{fig:f5} shows the color map
of $\delta a$ on the $\sigma_{\rm p}-\sigma_Z$ plane.  For an MBH with
$(a, i, \epsilon) = (0.99, 49\arcdeg, 125\arcdeg)$, as seen from this
panel, the MBH spin can be constrained to an accuracy of $\delta
a\sim0.1$ over a period of $\sim 45$\,yr if $(\sigma_{\rm p}, \sigma_Z)
= (10\microas, 1\kms)$. If the position and the redshift precisions can
be improved by a factor of a few, then the GC MBH spin may be
constrained with an even higher accuracy within the same time period or within
a relatively shorter time period (e.g., a decade) with the same accuracy.

We also generate mock observations for the hypothetical star over three full
orbits and use the MCMC technique to obtain the constraint on the MBH spin with
the mock observations. This star has a semimajor axis of $300\AU$, an
eccentricity of $0.95$ and an orbital orientation the same as that of S0-102.
The right panel of Figure~\ref{fig:f5} shows the dependence of the quality of
the spin constraint $\delta a$ on the position and the redshift precisions
obtained from the mock observations of this star. For an MBH with
$(a, i, \epsilon) = (0.99, 49\arcdeg, 125\arcdeg)$, as seen from this panel,
the MBH spin can be constrained to an accuracy of $\delta a \la 0.02$ over a
period of $\sim 10\yr$ if $(\sigma_{\rm p}, \sigma_Z) = (10\microas, 1\kms)$.
Since the position and the redshift differences induced by the precession due to
the quadrupole moment are about one percent to ten percent of the total
spin-induced signals, this star, if existing, can be used to put a constraint on
the quadrupole moment and thus possibly test the `no-hair' theorem. 
Note that \citet{W08} argued that at least two stars are needed to disentangle
the spin and the quadrupole moment, as the difference in the orbital
configurations of the two stars can help to break some degeneracy and determine
the involved parameters. Here even a single star can do this, because the
orbital configuration of the star evolves with time and the evolution effects
are incorporated into the time series of the mock observations, given the
sufficiently high measurement precisions and a long observational period.
Similarly, it was proposed that the spin and the quadrupole moment of the
central MBH can also be measured simultaneously by one pulsar given a high
timing precision \citep{Liuetal12}.

Although Figure~\ref{fig:f5} is obtained by setting the spin $a=0.99$, it can
be applied to other spin values if the spin has the same direction as that in
Figure~\ref{fig:f5}.  (For the region with $\delta a\ga a$, the MBH spin cannot
be constrained distinctively from the case with zero spin.) If the spin
direction is different from the one with the maximum spin-induced effects, the
spin-induced position displacement and redshift difference may be a factor of
$D$ and $F$ ($D,F>1$) smaller than the maximum ones, where the factors $D$ and
$F$ can be derived from
Figures~\ref{fig:f2} and \ref{fig:f3}. Therefore, the map for the quality of
the spin constraint can also be roughly obtained by replacing the figure labels
($\sigma_{\rm p}$, and $\sigma_Z$) in both panels of Figure~\ref{fig:f5} by
$D\sigma_{\rm p}$ and $F\sigma_Z$, respectively. We have checked this
numerically and find that it is appropriate. 

\subsubsection{Dependence on $(a_{\rm orb},e_{\rm orb})$}

The spin-induced signals depend on the semimajor axis and the eccentricity of
the star. The left panel of Figure~\ref{fig:f6} shows the quality of the spin
constraints by using the motion of three full orbits of hypothetical stars with
the same orbital orientation as that of S2/S0-2 but with various semimajor axes
and eccentricities. The input spin and its direction are
$(a, i, \epsilon)=(0.99, 49\arcdeg, 125\arcdeg)$ and the position and the
redshift precisions are $(\sigma_{\rm p}, \sigma_Z) = (10\mu{\rm
as}, 1\kms)$. Three orbital periods of a star with $a_{\rm orb}\simeq 300$~AU
are about 10~yr.  As shown in the figure, the spin and its direction can be
accurately constrained ($\delta a\la0.02$) within a short period (e.g., a
decade) by using the motion of a star with $a_{\rm orb} \la 300\AU$ and $e_{\rm
orb} \ga 0.95$.  In this case, a constraint on the quadrupole moment is also
possible since the contribution from the quadrupole moment can be as high as a
few to ten percent of the total spin-induced position displacements and
redshift differences.  With such a star and the assumed position and redshift
precisions $(10\mu{\rm as}, 1\kms)$, the MBH spin can still be well constrained
if the spin is not pointing close to the direction with the least spin-induced
effects, according to the difference in the signals with different spin
directions as shown in Figure~\ref{fig:f3}.  However, a star with $a_{\rm orb}
\ga 300\AU$ and $e_{\rm orb}\la 0.6$ may be not so useful in constraining the
GC MBH spin within a relatively short period (e.g., a decade), according to the
left panel of Figure~\ref{fig:f6}.

The right panel of Figure~\ref{fig:f6} shows the quality of the spin
constraints by using the motion of three full orbits of hypothetical stars with
the same orbital orientation as that of S0-102 and assuming the position and
the redshift precisions of $(\sigma_{\rm p}, \sigma_Z)=(50\microas, 5\kms)$.  The
input spin and its direction are $(a, i, \epsilon) =
(0.99, 49\arcdeg, 125\arcdeg)$. For such a spin direction, the spin-induced
effects are moderate, though not maximum, as illustrated in
Figure~\ref{fig:f3}.  As shown in the right panel of Figure~\ref{fig:f6}, the
spin can be constrained to an accuracy of $\delta a \la 0.1-0.3$ by using the
motion over a few orbits of a star with $a_{\rm orb} \la 300\AU$ and $e_{\rm
orb} \ga 0.95$ even if the astrometric and the velocity precisions are only about
$(50\microas, 5\kms)$.  Furthermore, for more than half of the possible spin
directions, the MBH spin can still be constrained by monitoring the motion of a
few orbits of such a star with the assumed position and redshift precisions of
$(50\microas, 5\kms)$.

The spin-induced effects also depend on the orbital orientation of a star
rotating around the MBH, and this dependence is partly incorporated into the
analytic approximations described by Equations~(\ref{eq:position}) and
(\ref{eq:redshift})-(\ref{eq:deltazv}).  By applying those analytical
approximations to the motion over a few orbits of an assumed star with
semimajor axis $\la 300\AU$, eccentricity $\ga 0.95$, but arbitrary orbital
orientations, we find that the GC MBH spin can be well constrained for most of
the possible spin directions if $(\sigma_{\rm p}, \sigma_Z)=(10\mu{\rm
as}, 1\kms)$. 

\begin{figure}
\centering
\subfigure[]{\includegraphics[scale=0.40]{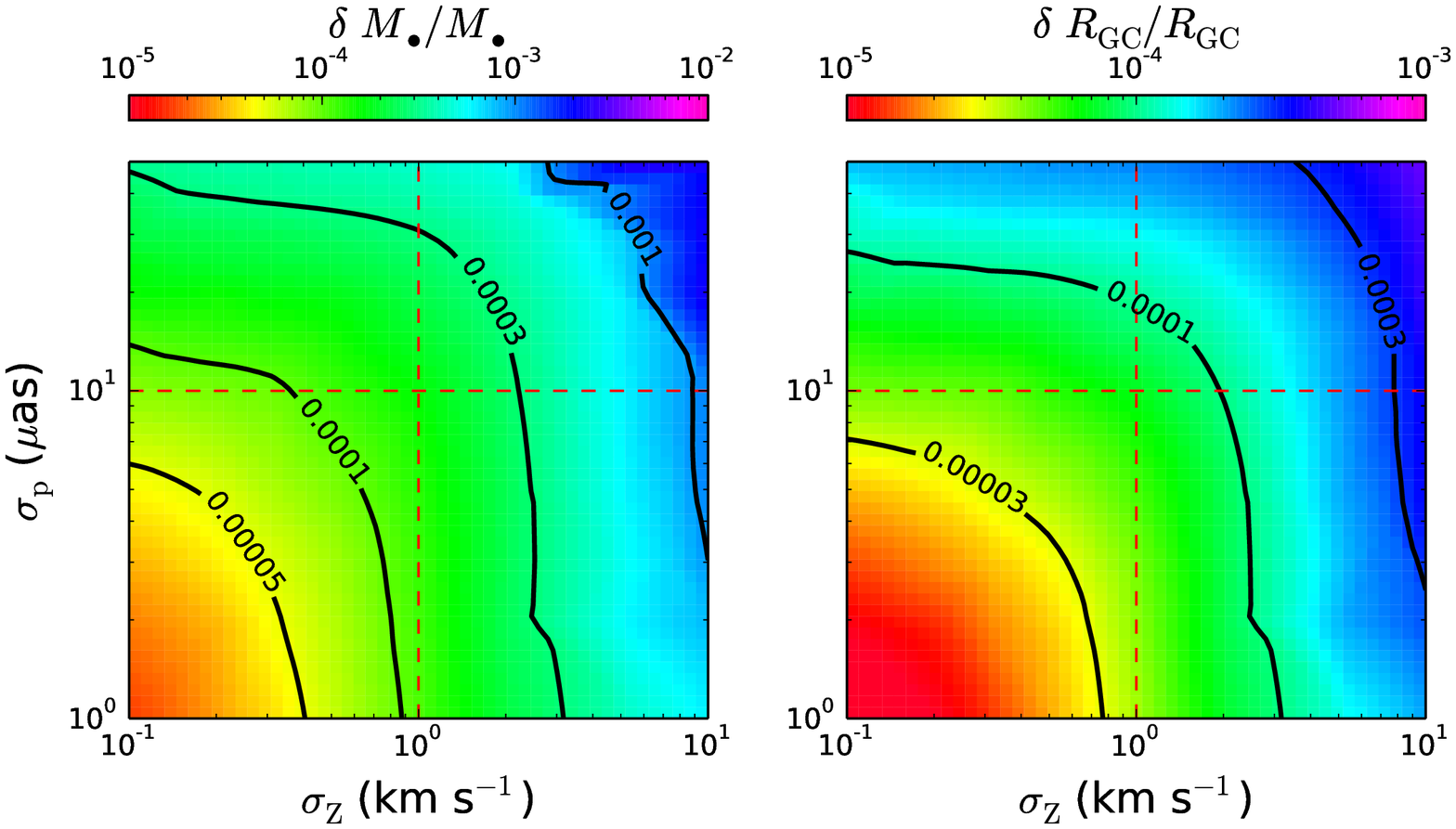}}
\subfigure[]{\includegraphics[scale=0.40]{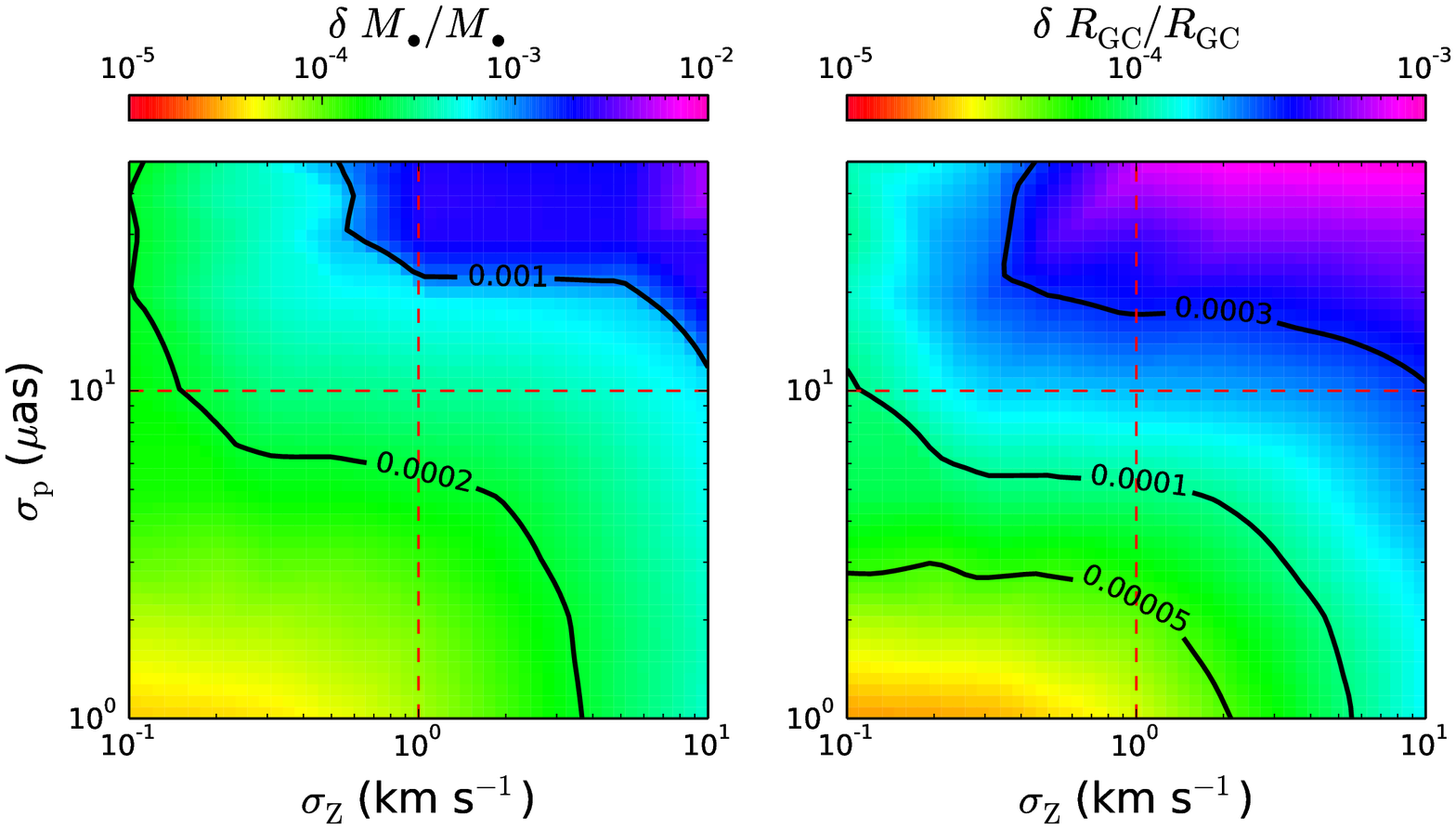}}
\caption{
Dependence of the qualities of the MBH mass constraint ($\delta \bh/\bh$)
and the GC distance constraint ($\delta R_{\rm GC}/R_{\rm GC}$) on the 
astrometric precision ($\sigma_{\rm p}$) and the  velocity precision 
($\sigma_Z$). The top panels show the results obtained from the mock 
observations of S0-2/S2 by assuming $(a, i, \epsilon)=(0.99, 49\arcdeg,
125\arcdeg)$; and the bottom panels show the results obtained from the 
mock observations for a hypothetical star with an orientation that is the same
as that of S0-102,
i.e., $(a, i, \epsilon)=
(0.99, 49\arcdeg, 125\arcdeg)$. The color bars indicate the values of 
$\delta \bh/ \bh$ or $\delta R_{\rm GC}/ R_{\rm GC}$, as labeled. The 
black lines represent the contours of $\delta \bh/ \bh$ or $\delta R_{\rm GC}
/ R_{\rm GC}$ with the marked values.
The figure indicates that the smaller semimajor axis
and the higher eccentricity of a star does not mean a higher-quality constraint
on the MBH mass and the GC distance.
In addition, although the improvement in velocity precision in the displayed
range is not effective at improving the quality of constraining the MBH spin
as shown in
Figure~\ref{fig:f5}, it can be effective at improving the quality of
constraining the MBH mass and the GC distance.
See Section~\ref{sec:massanddistance}.
}
\label{fig:f7}
\end{figure}

\subsubsection{Constraints on the MBH mass and the GC distance }\label{sec:massanddistance}

The qualities of the constraints on the MBH mass ($\bh$) and the GC distance
($R_{\rm GC}$) may be defined as $\delta \bh/ \bh$ and $\delta R_{\rm GC} /
R_{\rm GC}$, respectively, where $\delta \bh$ and $\delta R_{\rm GC}$ are the
$1\sigma$ errors obtained from the MCMC fittings to the mock observations.
Figure~\ref{fig:f7} shows  the qualities of the MBH mass constraint (left
panels) and the GC distance constraint (right panels), respectively, obtained
from the mock observations for S0-2/S2 (top panels) and a hypothetical star
(bottom panels) for a large range of the position and redshift accuracies.  As
shown in this figure, the accuracies of the MBH mass  and the GC distance
constraints can be up to a few $10^{-4}$ and $10^{-4}$, respectively, by using
the motion of S0-2/S2 over a few orbits if the astrometric and the velocity
accuracy of future facilities can reach $(\sigma_{\rm p}, \sigma_Z) =(10\mu{\rm
as}, 1\kms)$ as expected. Using a star with a significantly smaller semimajor
axis compared to S0-2/S2, e.g., $a_{\rm orb}=300\AU$, the accuracies of the MBH
mass and the GC distance constraints are poorer than those obtained by using
S0-2/S2 with the same $\sigma_{\rm p}$ and $\sigma_Z$. The main reason is that
the relative position error  is higher for a star with a smaller semimajor axis
given the same astrometric precision. Even if the astrometric precision and the
velocity precision are $(\sigma_{\rm p}, \sigma_Z) = (50\mu{\rm as}, 5\kms)$,
much poorer than the optimistic expectations for the future facilities, the
accuracies of the MBH mass and the GC distance constraints can still be up to
one thousandth and a few $10^{-4}$, respectively, by using the motion of
S0-2/S2 over three full orbits, although the MBH spin cannot be well
constrained for this case.

\section{Summary and Discussions}
\label{sec:summary}

In this paper, we adopt a full GR numerical method to investigate possible
constraints on the GC MBH spin that may be obtained by future facilities via
the relativistic motion of a star surrounding the MBH.  In the mapping of the
dependence of the spin-induced signals on any spin direction of the MBH for the
given example stars, the maximum and the minimum strength of the spin-induced
signals can differ by up to two orders of magnitude, and the strength of the
spin-induced signals are at the same order of magnitude as the maximum strength
if the spin direction is the same as the normal of the clockwise rotating young
stellar disk located at the sub-parsec scale of the GC. The motion of the stars
is helpful to test whether those two directions are the same or significantly
different and further provide insights into the assembly history of the MBH.
We use the MCMC fittings to demonstrate the quality of
constraining the MBH spin, given any set of the astrometric and the redshift
precisions of observational facilities.  Future facilities, such as the
GRAVITY on the VLTI, the TMT, and the E-ELT, are
expected to realize position and redshift (or velocity) precisions at the level
of $10-50\microas$ and $1\kms$, respectively.  We find that in the ranges of
astrometric precision and velocity precision with 1--30 $\mu$as and
0.1--10$\kms$, an improvement in astrometric precision would be more effective
at improving the quality of constraining the spin than an improvement in
velocity precision, although the improvement in velocity precision can be
effective at improving the quality of constraining the MBH mass and the GC
distance.  We obtain the parameter space of the semimajor axis and the
eccentricity for the orbit of the target star that a high-precision constraint
on the GC MBH spin can be obtained via the motion of the star.  We demonstrate
that the GC MBH spin can be constrained with high precision of 1-$\sigma$ error
$\la 0.1$ or even $\la 0.02$ by monitoring the motion of a star, if existing,
with a semimajor axis of $\la300\AU$ and an eccentricity of $\ga0.95$ in a
period of $\la10$\,yr with future facilities.  With such a star and the
optimistically expected position and redshift precisions, the quadrupole moment
of the MBH can also be constrained and thus the `no-hair' theorem may be tested
for optimistic cases in which the spin is pointing close to the
direction with the most significant spin-induced effects; a good constraint on
the MBH spin can still be obtained as long as
the spin direction is not close to that with the least significant spin-induced
effects.

Note here that the questions we addressed in this paper are based on some
significant simplifications; therefore, they should be taken as the most
optimistic results. In reality, there are a number of complications, such as,
possible perturbation from background stars and tidal dissipation of the target
stars.  It has been shown that the tidal dissipation of the target stars is
negligible for the parameter space studied in our paper
\citep{Psaltis13,PWK15}.  In the case in which the perturbation from background
stars is not negligible and the orbit of the target star has some additional
precession from the background stars, this precession may be modeled or removed
since its evolutionary pattern is different from that of the spin-induced effects,
which will be further studied in a separate work by using a hybrid relativistic
model.

\acknowledgments
This work was supported in part by the National Natural Science
Foundation of China under nos.\,11273004,\,11373031,\,and\,11390372,
and the Strategic Priority Research Program ``The
Emergence of Cosmological Structures'' of the Chinese Academy of
Sciences, Grant No. XDB09000000.

\end{document}